\documentstyle[aps,epsf]{revtex}

\newlength{\singsp}
\newcommand{\dspace}{\setlength{\baselineskip}{2\singsp}}

\def \MSbar {\vbox{\hrule\kern 1pt\hbox{\rm MS}}}

\begin{document}
%
\rightline{VAND-TH-97-03}
\rightline{AZPH-TH/97-04}
\bigskip
\centerline{
An Attempt to Determine the Largest Scale of Primordial Density
Perturbations in the Universe}
\bigskip
\centerline{ Arjun Berera$^{*}$, Li-Zhi Fang$^{**}$ and Gary Hinshaw$^{***}$}
\smallskip
\centerline{\it
$^{*}$Department of Physics and Astronomy,
Vanderbilt University,
Nashville, Tennessee 37235}
\vskip 0.1cm
\centerline{\it
$^{**}$Department of Physics,
University of Arizona,
Tucson, Arizona 85721}
\vskip 0.1cm
\centerline{\it
$^{***}$Laboratory for Astronomy and Solar Physics,
Code 685, NASA/GSFC,
Greenbelt, Maryland,  20771
}
\bigskip
\centerline{ }
\smallskip
\begin{abstract}
Causality constraints require a super-Hubble scale suppression in the
primordial power spectrum. This modification is implemented and
a three parameter likelihood analysis is performed of the
COBE-DMR 4-year data with respect to the amplitude, spectral index, and
suppression scale.  All suppression length scales larger than
$c/H_0$ are
consistent with the data, but scales of order $4c/H_0$ are slightly
preferred, at roughly the one-sigma level.
Many non-inflation models would be consistent with
a small suppression length scale, whereas for standard inflation models,
it would require small e-folds.
Suppression scales smaller than $c/H_0$ are strongly excluded
by the anisotropy data.
\end{abstract}

PACS numbers: 98.80.Es
\medskip

February 19, 1997

\medskip

\medskip

astro-ph/9703020

\narrowtext
%


Causality prohibits coherence between physical phenomena with
superhorizon scale separation.  For density perturbations which are
generated by local causal processes in a Robertson-Walker universe,
the implications of causality have been shown to imply a suppression
of the power spectrum which decreases faster than $k^4$ for scales
larger than the horizon \cite{abbtra}.  In non-inflationary cosmology
the horizon and Hubble radius are about the same size, so that no
causal mechanism could produce super-Hubble scale perturbations.
Inflationary cosmology is characterized by a time period in which the
horizon grew exponentially fast while the Hubble radius remained
essentially constant.  Thus inflation provides the only known
causal mechanism from present theory that allows
density perturbations to be super-Hubble scale with
respect to the present-day Hubble radius
$c/H_0 \approx 3000 \ {\rm h}^{-1} \ {\rm Mpc}$, where
${\rm h}^{-1} = 100/H_0 \ {\rm km} \ {\rm sec}^{-1} \ {\rm Mpc}^{-1}$.

For large scale structure, the power spectrum of the primordial
scalar density perturbations can be written as
\begin{equation}
P(k)= V Ak^n f(k) ,
\label{powspec}
\end{equation}
where $V$ is a large rectangular volume and
$f(k)$ is the long wavelength suppression factor, which
has been added to satisfy causality constraints.
The modification of a suppression scale to the pure power-law
power spectrum is mandated by causality.  Under very general
conditions, the suppression factor $f(k)$ can be obtained from causality
to be
\begin{equation}
f(k) = \frac{1}{1+(k_{min}/k)^m},
\label{fsup}
\end{equation}
where $k_{min}$ is the wavenumber of the suppression scale.
Causality places the
the strict constraint
for the suppression index
$m \geq 4-n$.  A suppression factor like eq. (\ref{fsup})
also has been
found in a model with cosmic strings plus cold or hot dark
matter \cite{albsteb}.  Eq. (\ref{powspec}) contains three parameters:
the power spectrum amplitude $A$,
the power-law spectral index n,
and the super-Hubble suppression scale wavenumber $k_{min}$.

There is a second constraint inferable from causality, which
is convenient from the
point of view of inflationary cosmology.  Models of primordial
density perturbations are generally classified as either inflationary
or non-inflationary. For both types, the long wavelength suppression
of the spectrum is a general
characteristic which is imposed by causality, although it is
typically ignored for inflation models.  Since for non-inflation
models the causal horizon is about the same as the present-day Hubble
radius, causality constraints also imply that
$k_{min} \sim \pi H_0$ in eq. (\ref{fsup}).
It should be noted that for non-inflation models, density
perturbations can be both primordial and generated after last
scattering,  whereas eq. (\ref{powspec}) strictly is valid for the
former type.  For many of the latter cases, eq. (\ref{powspec}) may still
be valid based on \cite{jaffe}.   The focus here is not on specific models,
although this caveat is worth stating.
For inflation models $k_{min}$ is far less constrained.  For
such models the largest scale of primordial density perturbations arose
from the fluctuations which first crossed the Hubble radius during
inflation \cite{kotu}.  The factor of expansion for them is given
by the parameter $N$, which is the number of e-folds of cosmic
scale growth.  In order to solve the horizon problem, inflation's
key purpose, the minimum required expansion places a lower limit
of $N \ge 50 \ - \ 70$.
Models of the standard inflationary cosmology offer no
convincing reason for $N$ to be near its lower limit and  generally
predict it to be several orders of magnitude bigger.  Hence,
for standard inflation models the super-Hubble
suppression length scale generically is expected
to be large,
$k_{min} \sim 0$.

Previous analyses of the COBE-DMR data have tacitly assumed that the
fundamental parameter $k_{min}$ was zero.  In this Letter we perform a maximum
likelihood fit of the 4-year DMR data with the three parameter spectrum in
eq. (\ref{powspec}).  We fix the suppression index $m$ at various values
and determine the most likely values of the amplitude A,
and spectral index n, in the presence of a third fundamental parameter, the
super-Hubble suppression scale wavenumber,
$k_{min}$, and we place limits on the
allowable range of $k_{min}$.

This analysis will consider the standard case of a
flat universe $\Omega_0=1$ with zero cosmological constant
$\Lambda = 0$.  The CMBR temperature fluctuation along the
direction unit vector ${\hat {\bf n}}$ is \cite{peebles}
\begin{equation}
\frac{\delta T ({\hat {\bf n}})}{T} \equiv
\frac{T({\hat {\bf n}}) -T}{T} = -\frac{H_0^2}{2 V c^2 }
\sum_{\bf k} \frac{\delta ({\bf k})}{k^2} e^{-i {\bf k} \cdot {\bf y}},
\label{delt}
\end{equation}
where $\delta({\bf k})$ is the Fourier amplitude of the density contrast
$\delta({\bf r})$ and
${\bf y}$ is a vector
of length $y = 2c/H_0$, the distance to the particle horizon in a matter
dominated universe,
which points in the direction
${\hat {\bf n}}$.
The power spectrum is defined as
\begin{equation}
P(k) \delta_{{\bf k}{\bf k}'} \equiv
<\delta({\bf k}) \delta({\bf k}')> = <|\delta({\bf k})|^2>
\delta_{{\bf k}{\bf k}'},
\end{equation}
where the brackets denote ensemble
average. $P(k)$ will be taken as eqs. (\ref{powspec}) and (\ref{fsup}).

The CMBR temperature fluctuations on the celestial sphere are usually
expressed by spherical harmonics
\begin{equation}
\delta T ({\hat {\bf n}})/T  = \sum_{lm} a_{lm} Y_{lm}({\hat {\bf n}}),
\end{equation}
where $Y_{lm}({\hat {\bf n}})$ are the spherical harmonic functions.
Defining a rotationally invariant coefficient
$C_l \equiv 1/(2l+1) \sum_m <|a_{lm}|^2>$, one finds from
eqs. (\ref{delt})
\begin{equation}
C_l = \frac{H_0^4}{2 \pi V c^4} \int_0^{\infty} dk \frac{P(k)}{k^2}
|j_l(ky)|^2 ,
\label{cl1}
\end{equation}
where $j_l(x)$ are the spherical Bessel functions.
Using eq. (\ref{powspec}), eq. (\ref{cl1}) becomes
\begin{equation}
C_l = \frac{A H_0^4}{2 \pi c^4 } \int_0^{\infty} dk
\frac{k^{n-2} |j_l(ky)|^2}{1+(k_{min}/k)^m}.
\label{pscoef}
\end{equation}
By the above conventions, the quadrapole anisotropy is given
as $Q_{rms-PS} =\sqrt{5C_2/4 \pi} T$ where $T=2.728K$
is the mean CMBR temperature \cite{cmbrtemp}.

To simplify this initial analysis, we have not considered temperature
fluctuations produced by tensor (gravitational) perturbations.
Most theoretical models expect the CMBR anisotropies to be dominated by
scalar perturbations.  Tensor perturbation will be subject to
a long wavelength suppression like eq. (\ref{fsup}), although
the power spectrum index $n$ generally should be different.

To check for possible
confusion from secondary sources of anisotropy, such as the Integrated
Sachs-Wolfe (ISW) effect, we have generated fully processed power spectra with
the code CMBFAST \cite{seza} for a range of the cosmological
parameters $\Omega_{vac}$ and $H$, for which
a significant ISW effect is possible.
The ISW effect generically increases the power in low-order multipoles and can
thus "fill in" some of the suppression generated by a cutoff scale.  However,
this requires values of $\Omega_{vac}$ in excess
of $ \sim 0.5$ to be significant, and in
no case can the ISW effect {\it mimic} the effects of a
small suppression length scale.
Thus we do not further consider the ISW effect in this paper since it cannot
{\it decrease} the significance of any possible detection of a cutoff scale.
An alternative possibility is that an anti-alignment of the
Galaxy quadrupole with the CMBR quadrupole would suppress the quadrupole power.
However, Galaxy modeling indicates that this is not a significant effect
\cite{kogut}.

To fit the parameters of our model power spectra, we will use the pixel based
likelihood method (pixed based method)
introduced in \cite{bond} and used in COBE-DMR studies
\cite{tebu,ghin}.  This method is predated by Gaussian likelihood fits to
the 2-point angular correlation function (2-point method).  The
primary disadvantage of the 2-point method is that the 2-point
correlation function is not Gaussian distributed, so the Gaussian
likelihood fit is only approximate.

In the pixel based method the covariance matrix is computed between map
pixels i and j as
\begin{equation}
M_{ij} \equiv <\frac{\delta T_i}{T} \frac{\delta T_j}{T}>
= \frac{1}{4 \pi} \sum_l (2l+1) W_l^2 C_l
P_l({\hat n}_i \cdot {\hat n}_j) ,
\end{equation}
where $T_i$ is the temperature in pixel $i$ of a map, $W_l^2$ is
the experimental window function that includes the effects of beam
smoothening and finite pixel size, $C_l$ is given in eq. (\ref{pscoef}),
$P_l({\hat n}_i \cdot {\hat n}_j)$ is the Legendre polynomial
of order $l$, and ${\hat n}_i$ is the unit vector towards the center of
pixel $i$.  For pixel temperatures that are Gaussian distributed, the
covariance matrix fully specifies the statics of the temperature
fluctuations.  The probability of observing a map with pixel
temperatures $\vec{T}$, given a model $C_l$, is
\begin{equation}
P(\vec{T}|C_l({\bf p})) d{\vec T} = \frac{d {\vec T}}{(2 \pi )^{J/2}}
\frac{e^{-\frac{1}{2} {\vec T}^T \cdot M^{-1}(C_l({\bf p})) \cdot
{\vec T}}}{\sqrt{{\rm det} M(C_l({\bf p}))}}
\end{equation}
where $J$ is the number of pixels in the map.  Assuming a uniform prior
distribution of cosmological model parameters, the probability,
or likelihood function, of a given $C_l$ with parameters
${\bf p}$ and a given map ${\vec T}$ is then
\begin{equation}
L(C_l({\bf p})|{\vec T}) \propto
\frac{e^{-\frac{1}{2} {\vec T}^T \cdot M^{-1}(C_l({\bf p})) \cdot
{\vec T}}}{\sqrt{{\rm det} M(C_l({\bf p}))}} .
\end{equation}
For convenience we will denote the likelihood function
simply as $L({\bf p})$.

We have evaluated the above likelihood function using the model
power spectra
in eq. (\ref{pscoef}) for two cases: sharp cutoff $m=\infty$ and
minimal cutoff $m=4-n$. The results reported below are from the
COBE correlation technique map,
which has the best estimate of the high-latitude Galaxy
subtracted off \cite{kogut}.
We have tested a case with a map which has no residual
Galaxy subtracted and the results are not qualitatively different.

Full details of our tri-parameter likelihood analysis will be reported
elsewhere.  Here we will present our results in terms of
the projected likelihoods $L(k_{min},n;Q_{rms-PS}),
L(k_{min};Q_{rms-PS},n)$, and $L(n;k_{min},Q_{rms-PS})$.
For a likelihood function $L({\bf p})$, defined
by a set of parameters ${\bf p}=({\bf p}_1, {\bf p}_2)$, the
projected likelihood $L({\bf p}_1;{\bf p}_2)$ is defined as
$L({\bf p})$ for fixed ${\bf p}_1$ evaluated at the most likely
${\bf p}_2$.  For $L(k_{min},n;Q_{rms-PS})$ the most likely
$(k_{min},n)$ are given in table 1 with $68 \%$ confidence level (CL)
uncertainties
with respect to the projected likelihoods $L(k_{min};n,Q_{rms-PS})$ and
$L(n;k_{min},Q_{rms-PS})$.  Table 2 gives the projected likelihood results for
both cutoff models for $L(k_{min},Q_{rms-PS})$ under the constraint $n=1$.
In both tables 1 and 2, the most likely quadrapole anisotropies
$Q_{rms-PS}$, are given with $68 \%$ CL uncertainties from
the likelihood evaluated at the most likely values of $k_{min}$
and $n$ (or fixed $n=1$ for table 2).
These errors reflect the precision of the normalization
for the specified models for a fixed shape of the
spectrum ie. ($k_{min},n$).
Finally table 3 gives the $99 \% $ confidence upper limits on $k_{min}$
for unconstrained $n$ and $n=1$.  Our results reproduce those
given in \cite{ghin,fouryear} in the pure power-law
limit, $k_{min} = 0$.

Table 3 confirms a detection of a coherence length
bigger than the Hubble radius, $c/H_0$, for both cutoff models.
Tables 1 and 2 show that all values of the suppression length
scale larger than the
Hubble diameter ($\lambda_{max} \sim 2c/H_0$) are consistent with data,
although length scales of order $4c/H_0$ are slightly preferred.
If the spectral index is fixed at $n=1$ (Table 2),
there is a one sigma exclusion
of $k_{min}=0$.  However with $n$ left unconstrained (Table 1), only the sharp
cutoff model is found to exclude $k_{min} = 0$ at $68 \%$ confidence (Fig. 1).
In no case was a two sigma exclusion of $k_{min}=0$ found.
The quadrapole anisotropies $Q_{rms-PS}$ in all cases in tables 1 and 2
are smaller than the pure power-law result of
$Q_{rms-PS}= 15.3^{+3.8}_{-2.8} \ \mu K$ \cite{ghin,fouryear}, because
of the low-$\ell$ suppression in the spectrum.  Note that our
results for $Q_{rms-PS}$ are comparable to those in \cite{ghin}
where the quadrapole $C_2$ was fit independent of the rest of the
spectrum.  This suggests that
the shape of
the likelihood in our current analysis is being driven primarily by the low
quadrupole, and the most-likely normalized spectra are such that the mean
quadrupole in each case is comparable to the actual quadrupole in our sky.

One can get a feeling for the relative preference between the
sharp and minimal cutoffs by examining the likelihood at $k_{min}=0$,
where both model spectra are common.  By this approach the sharp cutoff
is slightly preferred to the minimal cutoff by a factor $1.24$.
Note that the most-likely spectral index $n$ is closer to the scale-invariant
limit ($n=1$) with non-zero $k_{min}$ than it was with $k_{min}=0$ ($n=1.2
\pm 0.3$ at $k_{min}=0$ \cite{ghin,fouryear}).
In summary, a finite super-Hubble
suppression length scale is, at best, suggested by the data.

Two features of the power spectrum model should
be mentioned.  First, for very large
suppression length scales, $k_{min}y < 1$, the model spectra are virtually
indistinguishable from a pure power-law spectrum.  Therefore this analysis is
most powerful at placing upper limits on the suppression scale
wavenumber $k_{min}$.  Second,
there is some degeneracy between $k_{min}$ and $n$, particularly for the
minimal cutoff model with $m = 4-n$: decreasing
the suppression scale $k_{min}$
steepens the slope of the power spectrum at low spherical harmonic order
$\ell$, which partially mimics the effect of increasing the spectral index
$n$.  Additional large and medium scale anisotropy data, as expected from the
forthcoming satellites MAP and Planck, should allow us to place better limits
on $n$ so as to partly constrain this degeneracy.

To cross-check the results of our likelihood analysis, we simulated
$1000$ pure power-law, scale-invariant skies, $(k_{min},n)=(0.0,1.0)$,
to determined what fraction have
likelihood functions similar to the data.  For the upper limit on $k_{min}$,
the Monte Carlo results firmly confirm the $99 \% $ CL
given in table 3 for both types of cutoffs.  For the lower limit on $k_{min}$,
the Monte Carlo results indicate that a pure power-law, scale invariant
universe has a $20 \%$ ($33 \% $)
chance of spuriously imitating
a universe with a super-Hubble suppression scale of the size favored
by the data with a sharp (minimal) cutoff.
These results are consistent with the likelihood analysis.
In particular they verify that the tendency found in the data for a
small suppression length scale is not an effect of anomalously large spurious
fluctuations.
The Monte Carlo analysis also indicates that increasing the signal-to-noise
ratio with further data may give a 1-2 sigma greater discrimination of
$k_{min}=0$, but cosmic variance prohibits much greater significance.  Thus,
while the evidence for a finite suppression scale is not statistically
significant, we find it suggestive enough to consider its implications.

The standard inflation models
predict $k_{min} \sim 0$, which is consistent with, but not
preferred by, the data.
It is worth noting that the super-Hubble
suppression behavior in the basic new inflation \cite{bst}
and chaotic inflation \cite{chainf} scenarios is closer to
the sharp cutoff form.
For a large class of non-inflationary models there is no mechanism for the
growth of super-Hubble scale perturbations.  In models with a "late time"
cosmological phase transition \cite{latet},  we expect
$k_{min} y \geq 4$ \cite{abbtra,jaffe}.  In models with topological defects,
in which the perturbations are produced before last-scattering, the
super-Hubble suppression scale depends on the dynamics of the defects. For
example, in models with cosmic strings plus hot or cold dark matter it was
found that $k_{min}y \sim 2.1 \ - \ 7.9$ with a cutoff behavior closer
to the minimal form \cite{albsteb}.   These two types of non-inflation
models are potentially consistent with the COBE data, but subhorizon
evolution must be checked.  For cosmic strings, recent simulations
\cite{allen} indicate that subhorizon evolution
induces greater power in the low-order multipoles, $C_\ell$.
If further study supports
this finding, it will imply that a small super-Hubble suppression
length scale in the
COBE data cannot be explained by cosmic strings.
Both the super-Hubble suppression scale and cutoff should be noted in
all theoretical models of primordial density perturbations.

If further investigation substantiates the suggestion of a
small super-Hubble suppression length scale, there are at least two possible
explanations.  One is that density perturbations
produced by non-inflation models are the dominant contributor.
The second is that density perturbations are primarily produced
during inflation, but that the number of e-folds, N, is
close to its lower bound.  In standard inflation models
a small $N$ is viewed as a fine tuning of the theory.  Nevertheless,
a small $N$ is consistent from two other directions.
Firstly, observationally a small $N$ also could consistently explain
a nearly-flat to open universe \cite{gfre}.  We
plan to study the suppression scale in low density power spectrum models.
Secondly,
it has been shown from standard Friedmann cosmology \cite{berera4}
that the
requirement of small $N$
can be realized by a symmetry breaking
phase transition at finite temperature, during which the universe
smoothly goes from an inflation-like stage to the radiation dominated
stage without an intermediate reheating period.  The naturalness of a
small $N$ found in this study \cite{berera4} motivated the search in
this paper for a small super-Hubble suppression length scale.
Scenarios in this non-isentrephic inflation-like expansion regime,
which was named in \cite{berera4} the big-bang-like inflation regime,
avoid requiring both
localized fields on ultraflat potential
surfaces and an impulsive large scale energy release
during a reheating period \cite{kotu}, which
must be isotropic over causally vastly-disconnected regimes. Both these
fundamental conundrums have been difficult to resolve in the
standard inflation picture. The theory of density perturbations
\cite{bf2} in this
intermediate regime of radiation and vacuum energy requires further
study of warm inflation \cite{wi} and possible other mechanisms
before direct comparison is possible with standard inflation models.
The increased complexity of this mixed fluid regime
poses several theoretical questions which must be understood before
firm predictions can be made.

The super-Hubble suppression scale added to the power spectrum in this paper
is required on first principle by any causal theory.  It is more
fundamental than the spectral index and the amplitude and much less
model-dependent than other processes that can
effect the large-scale anisotropy, such as the Integrated Sachs-Wolfe effect.
Moreover, the
suppressing effect of causality on the power spectrum can produce a
potentially significant effect on large scales that is not readily confused
with other processes that generically boost the large-scale anisotropy power.
This possibility calls for a reexamination of existing parameter fits to the
COBE data.

In conclusion, we modified the primordial power spectrum of density
fluctuations from a pure power-law form to a form that includes a
super-Hubble suppression scale, $k_{min}$, so as to properly respect
causality constraints.  We fit this spectrum to the 4-year COBE-DMR
data and find that the data prefers a finite suppression scale, but does not
rule out $k_{min}=0$.
The best fit to the COBE 4-year data is
$(n, k_{min}y, Q_{rms-PS}) =
(1.07^{+0.32}_{-0.35}, 3.4^{+0.7}_{-2.2}, 10.9 \pm 0.7 \mu K)$,
with a slight preference for a sharp super-Hubble cutoff.
Upper limits
on $k_{min} y$ have been firmly placed.
We conclude that this third fundamental parameter is
measurable from the COBE data.

Financial support was provided in part by
the U. S. Department of Energy and the NASA Office of Space Sciences.

FIGURE CAPTIONS

Figure 1: Relative likelihood for the suppression
scale wavenumber, $k_{min}$, projected over
$n$ and $Q_{rms-PS}$ with $y=2/H_0$.

\eject

\begin{center}
\begin{tabular}{c|ccc}
super-Hubble      & $k_{min} y$ & $n$ & $Q_{rms-PS}$ \\
 cutoff          &           &   & ($\mu K$) \\ \hline
Minimal ($m=4-n$)  & $2.9^{+1.8}_{-2.9}$ & $1.06^{+0.48}_{-0.67}$ & $13.0
\pm 0.9$  \\
\\
Sharp ($m= \infty$)  &  $3.4^{+0.7}_{-2.2}$ & $1.07^{+0.32}_{-0.35}$  & $10.9
\pm 0.7$ \\
\end{tabular}

\vspace{5 mm}

Table 1: Maximum likelihood parameter estimates for unconstrained
$n$ with $68 \% $ CL uncertainties from the respective projected
likelihood functions.

\vspace{20 mm}
\vspace{10 mm}

\begin{tabular}{l|cc}
super-Hubble     & $k_{min} y$            & $Q_{rms-PS}$ \\
  cutoff ($n=1$)                  &                        & ($\mu K$) \\
\hline
Minimal ($m=4-n$)  & $3.2^{+1.5}_{-2.0}$    &  $12.8 \pm 0.9$  \\
\\
Sharp ($m= \infty$)  &  $3.5^{+0.8}_{-1.8}$ & $10.8 \pm 0.7$ \\
\end{tabular}

\vspace{5 mm}

Table 2: Maximum likelihood parameter estimates for constrained
$n=1$ with $68 \% $ CL uncertainties from the respective projected
likelihood functions.

\vspace{20 mm}
\vspace{10 mm}
\begin{tabular}{l|cc}
super-Hubble cutoff       & $k_{min} y$       & $k_{min} y|_{(n=1)}$ \\ \hline
Minimal ($m=4-n$)  & $10.5$     &  $7.2$  \\
\\
Sharp ($m= \infty$)  &  $5.0$   & $4.9$ \\
\end{tabular}

\vspace{5 mm}

Table 3: $99 \% $ CL upper limits on $k_{min} y$.

\end{center}
\dspace


\begin{thebibliography}{99}
\bibitem{abbtra}  L. F. Abbott and J. Traschen, Astrophys. J. {\bf 302}.
39 (1986); J. Robinson and B. D. Wandelt, Phys. Rev. {\bf D53}, 618
(1996).

\bibitem{albsteb} A. Albrecht and A. Stebbins, Phys. Rev. Lett. {\bf 68},
2121; {\bf 69}, 2615 (1992).

\bibitem{jaffe} A. H. Jaffe, A. Stebbins, and J. A. Frieman, Astrophys.
J. {\bf 420}, 9 (1994).

\bibitem{kotu}
E.W.Kolb and M.S. Turner,
{\it The Early Universe}, (Addison-Wesley, New York, 1990).

\bibitem{peebles} P.J.E. Peebles, Astrophys. J. {\bf 263}, L1 (1982).

\bibitem{cmbrtemp} D, J. Fixsen et. al., Astrophys. J. {\bf 473},
576 (1996).

\bibitem{seza} U. Seljak and M. Zaldarriaga, Astrophys. J. {\bf 469},
437 (1996).

\bibitem{kogut} A. Kogut et. al., Astrophys. J. {\bf 464}, L5 (1996).

\bibitem{bond} J. R. Bond, Phys. Rev. Lett. {\bf 74}, 4369 (1995).

\bibitem{tebu} M. Tegmark and E. F. Bunn, Astrophys. J. {\bf 455}, 1
(1995).

\bibitem{ghin} G. Hinshaw et. al., Astrophys. J. {\bf 464}, L17 (1996).

\bibitem{fouryear} C. L. Bennett et. al., Astrophys. J. {\bf 464},
L1 (1996).

\bibitem{bst} J.M. Bardeen, P.J. Steinhardt, \& M.S. Turner,
Phys. Rev. {\bf D28}, 679 (1983);

\bibitem{chainf} A. Linde, Phys. Lett. {\bf 129B}, 177 (1983).

\bibitem{latet} I. Wasserman, Phys. Rev. Lett. {\bf 57}, 2234 (1986).

\bibitem{allen} B. Allen et. al., astro-ph/9609038.

\bibitem{gfre} G. F. R. Ellis, Class. Quantum Grav. {\bf 5},
891 (1988).

\bibitem{berera4} A. Berera, Phys. Rev. {\bf D55}, 3346 (1997).

\bibitem{bf2} A. Berera and L. Z. Fang, Phys. Rev. Lett. {\bf 74},
1912 (1995).

\bibitem{wi} A. Berera,  Phys. Rev. Lett. {\bf 75},
3218 (1995).

\end{thebibliography}
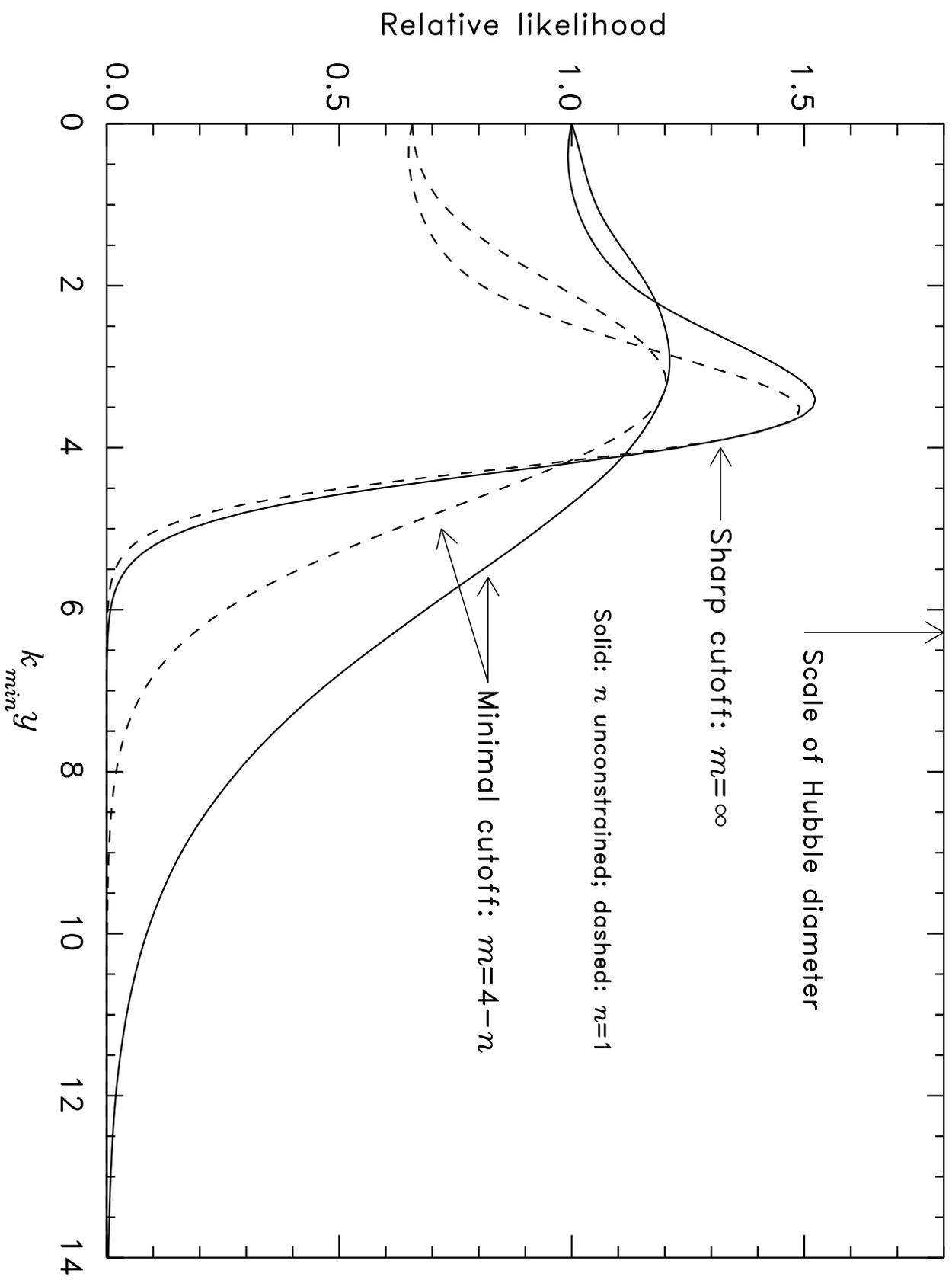
\end{document}